\documentclass[11pt,leqno,fleqn]{article}
\usepackage{amsmath}
\usepackage{amsfonts,amssymb,amsthm}
\usepackage{graphicx,color}
\usepackage{natbib}

\begin{document}
\newcommand{\ISItitle}[1]{\vskip 0pt\setlength{\parindent}{0cm}\Large\textbf{#1}\vskip 12pt}
\newcommand{\ISIsubtitleA}[1]{\normalsize\rm\setlength{\parindent}{0cm}\textbf{#1}\vskip 12pt}
\newcommand{\ISIsubtitleB}[1]{\normalsize\rm\setlength{\parindent}{0cm}\textbf{#1}\vskip 12pt}
\newcommand{\ISIsubtitleFig}[1]{\normalsize\rm\setlength{\parindent}{0cm}
\textbf{\textit{#1}}\vskip 12pt}
\newcommand{\ISIauthname}[1]{\normalsize\rm\setlength{\parindent}{0cm}#1 \\}
\newcommand{\ISIauthaddr}[1]{\normalsize\rm\setlength{\parindent}{0cm}\it #1 \vskip 12pt}

\newtheorem{theorem}{Theorem}
\newtheorem{proposition}[theorem]{Proposition}
\newtheorem{lemma}[theorem]{Lemma}
\newtheorem{corollary}[theorem]{Corollary}
\newtheorem{definition}[theorem]{Definition}
\theoremstyle{definition}
\newtheorem{remark}[theorem]{Remark}
\newtheorem{example}[theorem]{Example}

\ISItitle{Evolution of bacterial genomes under horizontal gene
  transfer}

\ISIauthname{Baumdicker, Franz} 
\ISIauthname{Pfaffelhuber, Peter} 
\ISIauthaddr{Albert-Ludwigs-Universit\"at Freiburg, Abteilung f\"ur Mathematische Stochastik\\
Eckerstra\ss e 1\\
79104 Freiburg, Germany\\ 
E-mail: baumdicker@stochastik.uni-freiburg.de\\
E-mail: p.p@stochastik.uni-freiburg.de}

\ISIsubtitleA{Introduction} %
Unraveling the evolutionary forces shaping
bacterial diversity can today be tackled using a growing amount of
genomic data. In recent years, the number of completely sequenced
prokaryotic genomes has increased to around 1700 (NCBI). In
particular, first datasets are available for samples of complete
genomes from closely related strains, which are of the same bacterial
species \citep{Medini2005,Tettelin2005,Tettelin2008}. Such datasets
mark a revolution of microbial evolutionary biology, which turned from
a theory-rich/data-poor subject into a data-rich/theory-poor field in
the last decade.

While the genome of eukaryotes is highly stable, bacterial genomes
from cells of the same species highly vary in gene content. For
example, the pathogenic strain \emph{E. coli} O157:H7 carries 1387
genes which are absent in the commensal \emph{E. coli} K-12
\citep{Perna2001}. This huge variation in gene content led to the
concepts of the \emph{distributed genome} of bacteria and their
\emph{pangenome} \citep{Tettelin2005,Ehrlich2005}. In datasets, genes
present in all genomes of a taxon are called \emph{core genes} while
genes present in only some but not all individuals comprise the
\emph{accessory genome}. 

Gene content diversity originates in horizontal exchange of genomic
material and pseudogenozation followed by gene loss. In particular,
the amount of genetic exchange within a bacterial species determines
the level of clonality \citep{MaynardSmithSmithRourkeSpratt1993}. 
There are three different mechanisms of horizontal genetic exchange:
(a) Transformation is the uptake of genetic material from the
environment. (b) When a bacterium is infected by a lysogenic virus (phage) it
provides additional genetic material that can be built in the
bacterial genome. This process is known as transduction. (c)
Conjugation requires a direct link (pilus) between two bacterial cells
and leads to exchange of genetic material. These three mechanisms are
usually referred to as horizontal gene flow. Recently, small virus-like elements
called Gene Transfer Agents (GTAs) have been hotly debated to be the most
important source for horizontal genetic exchange in some species
\citep{pmid20929803}.

We present a population genetic model for gene content evolution which
accounts for several mechanisms. Gene uptake from the environment is
modeled by events of \emph{gene gain} along the genealogical tree,
which describes the relationships between the individuals of the population.
Pseudogenization may lead to deletion of
genes and is incorporated by \emph{gene loss}. These two mechanisms
were studied by \cite{Huson:2004:Bioinformatics:15044248} using a
fixed phylogenetic tree. Taking the random genealogy given by the
coalescent \citep{Kingman1982,Hudson1983}, we studied the resulting
genomic diversity already in \cite{BaumdickerHessPfaffelhuber2010}
(see also \citealp{BaumdickerHessPfaffelhuber2011}). In the present
paper, we extend the model in order to incorporate events of
intraspecies horizontal gene transfer. Within this model, we derive
expectations for the gene frequency spectrum and other quantities of
interest.

\bigskip

\ISIsubtitleA{The model} %
We consider the following model for bacterial
evolution: Each bacterial cell carries a set of \emph{genes} and every
gene belongs either to the \emph{core genome} or the \emph{accessory
  genome}. The infinite set $I:=[0,1]$ is the set of conceivable
accessory genes and $\mathcal G_c$ with $\mathcal G_c \cap I =
\emptyset$ is the core genome. A population of constant size consists
of $N$ individuals (bacterial cells). We model the accessory genome of
individual $i$ by a finite counting measure $\mathcal G_i(t)$ on
$I$. We will identify finite counting measures with the set of atoms,
i.e.\ we write $u\in \mathcal G_i(t)$ if $\langle \mathcal G_i(t),
1_u\rangle \geq 1$.

The population evolves according to Wright--Fisher dynamics. That is,
generations are discrete and individual (bacterial cell) $j$ in
generation $t+1$ \emph{chooses} a parent from generation $t$ purely at
random and independent of all other individuals at time $t+1$. We
denote that parent by $A_j(t)$. In order to obtain the genome
$\mathcal G_j(t+1)$, we follow the mechanisms:

\begin{enumerate}
\item \emph{Gene loss}: Denote the $1-\rho/(2N)$-thinning of $\mathcal
  G_{A_j(t)}(t)$ by $\mathcal G'_j(t+1)$. That is, $u\in \mathcal
  G'_j(t+1)$ iff $u\in \mathcal G_{A_j(t)}(t)$ and an independent coin
  with success probability $1-\rho/(2N)$ shows a success.
\item \emph{Gene gain}: Choose an independent random counting measure
  $\mathcal H'_j(t+1)$ according to a Poisson process on $I$ with intensity
  $\theta/(2N)$. 
\item \emph{Horizontal gene transfer}: For every $i=1,\dots,N$ (the
  \emph{donor}) and $v\in \mathcal G_i(t)$ (the \emph{transferred
    gene}), let $v\in \mathcal H_j''(t+1)$ with probability
  $\gamma/(2N^3)$. In this event, individual $j$ is called the
  \emph{acceptor} of gene $v$.
\end{enumerate}
Finally, set
\begin{align}\notag
  \mathcal G_j(t+1) = (\mathcal G_j'(t+1) + \mathcal H_j'(t+1) + \mathcal
  H_j''(t+1))\wedge 1
\end{align}
for the genome of individual $j$ in generation $t+1$.  The '$\wedge
1$'-term indicates that we do not model paralogous genes, i.e.\
horizontal gene transfer events have no effect if the acceptor
individual $j$ already carries the transferred gene. \sloppy We refer
to $(\mathcal G_1(t), \dots, \mathcal G_N(t))_{t=0,1,2,\dots}$ undergoing
the above dynamics as the Wright--Fisher model for bacterial genomes
with horizontal gene flow. It can be shown that this Markov chain is
Harris recurrent and hence, has a unique equilibrium.

\begin{figure}
  \begin{center}
    \includegraphics[width=3in]{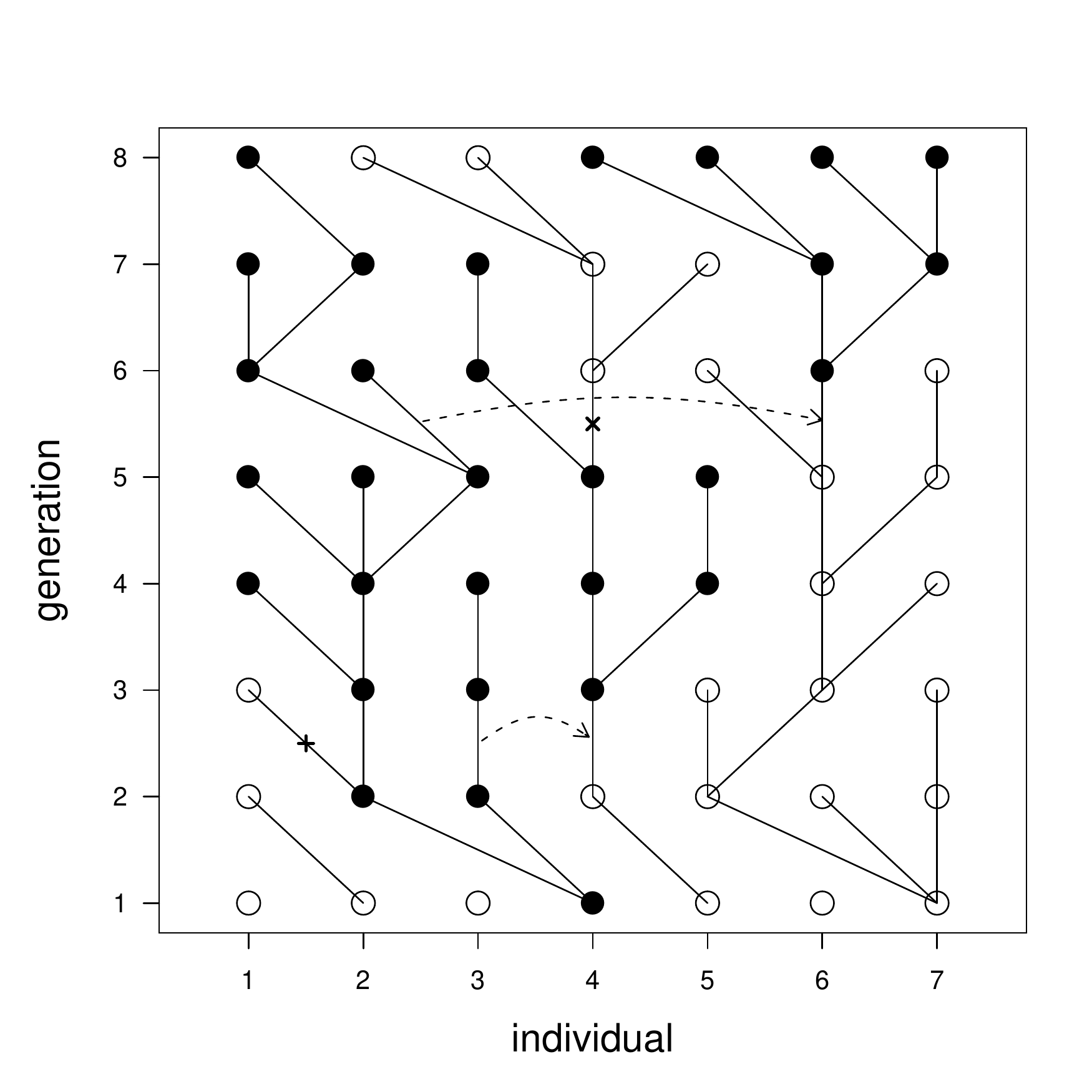}
    \caption{\label{fig3} Genes are with probability $\theta/(2N)$ per
      individual per generation. In this illustration only one gene is
      gained in generation 1 in individual 4.  Individuals carrying
      this gene are shown in black. Offspring inherits the gene from
      its ancestor, unless a loss event occurs (cross), with
      probability $\rho/(2N)$. With probability $\gamma/(2N^3)X(1-X)$,
      a gene is transferred to a random individual, such that now both
      the donor and the acceptor carry the gene.}
  \end{center}
\end{figure}

We are mainly interested in large populations and a rescaling of time
by a factor of $N$. The corresponding limit is usually refered to as
large population limit in the population genetic literature. The
following argument is crucial in the proof of our main result,
Theorem~\ref{T1}. Let $X^N(t)$ be the frequency of gene $u$ in
generation $[tN]$ in the population of size $N$. Then, in the large
population limit, $N\to\infty$, the process $(X^N(t))_{t\geq 0}$
converges weakly to the solution of the SDE
\begin{align}\label{eq:SDE}
  dX = \big(-\tfrac\rho 2 X + \tfrac \gamma 2 X(1-X) \big)dt +
  \sqrt{X(1-X)} dW
\end{align}
for some Brownian motion $W$. To see this, note that the evolution of
frequencies of gene $u$ is an autonomous process. The diffusion term
is associated with random reproduction events and has the well-known
form from~\eqref{eq:SDE} \citep{Ewens2004}, known as Wright--Fisher
noise. Gene loss reduces $X^N$ with probability approximately
proportional to $X$ and $\rho/(2N)$. After rescaling of time, this
turns into the rate $-(\rho X/2) dt$. Last, horizontal gene transfer
increases $X^N$ with probability approximately proportional to
$\gamma/(2N^3)$ and to the number of pairs where the horizontal gene
transfer events has an effect, $N^2 X^N(1-X^N)$. After rescaling of
time, this turns into the rate $\tfrac\gamma 2 X(1-X)dt$.

\bigskip

\ISIsubtitleA{Sample statistics} %
Consider a sample $\mathcal
G_1,\dots,\mathcal G_n$ of size $n$ taken from the population. We
consider several statistics under the above dynamics:

The \emph{average number of genes (in the accessory genome)} is given
by
\begin{align}\label{eq:Hbar}
  A := A^{(n)} := \dfrac 1n \sum_{i=1}^n |\mathcal G_i|
\end{align}
where $ |\mathcal G_i| := \langle \mathcal G_i, 1\rangle$ is the total
number of accessory genes in individual $i$.

The \emph{average number of pairwise differences} is given by
\begin{align}\label{eq:Dbar}
  D := D^{(n)} := \dfrac 1{n(n-1)} \sum_{1\leq i \neq j\leq n}
  |\mathcal G_i \setminus \mathcal G_j|
\end{align}
where $ \mathcal G_i \setminus \mathcal G_j := (\mathcal G_i -
\mathcal G_j)^+$ are the genes present in $i$ but not in $j$.

The \emph{size of the accessory genome} is given by
\begin{align}\label{eq:G}
  G:=G^{(n)}:=\Big| \bigcup_{i=1}^n \mathcal G_i \Big|
\end{align}
where $ \bigcup_{i=1}^n \mathcal G_i = \Big(\sum_{i=1}^n \mathcal
G_i\Big) \wedge 1$ is the set of genes present in any individual from
the sample.

The \emph{gene frequency spectrum (of the accessory genome)} is given
by $G_1:=G^{(n)}_1,\dots,G_n:=G^{(n)}_n$, where
\begin{align}\label{eq:Gk}
  G_k^{(n)} := G_k := |\{u\in I: u\in\mathcal G_i \text{ for exactly }
  k \text{ different }i\}|.
\end{align}

\bigskip

\ISIsubtitleA{Results} %
Using diffusion theory, we obtain first moments
of all of the above statistics in equilibrium. We start with
expectations of $G_1^{(n)},\dots,G_n^{(n)}$, since all other quantities
can be expressed in terms of the gene frequency spectrum. The proof of
Theorem~\ref{T1} can be found at the end of the manuscript.

\begin{theorem}[Gene frequency spectrum]
  \label{T1} Consider a sample of size $n$ taken from the
  Wright--Fisher model for bacterial genomes with horizontal gene flow
  with $\rho>0, \theta>0, \gamma\geq 0$ in equilibrium. Then, as
  $N\to\infty$,
  \begin{align}\label{eq:T1}
    \mathbb E[ G_k^{(n)} ] &= \frac{\theta}{k} \frac{n \cdots
      (n-k+1)}{(n-1+\rho) \cdots (n-k+\rho)} \Big(1 +
    \sum\limits_{m=1}^{\infty} \frac{(k)_m \gamma^m}{(n+\rho)_m m!}\Big)
  \end{align}
  with $ (a)_b := a (a+1) \cdots (a+b-1).$
\end{theorem}

\begin{corollary}[More summary statistics of gene content] \mbox{}\\
  Under \label{C1} the same assumptions as in Theorem~\ref{T1},
  \begin{align}
    \mathbb E[A^{(n)}] &= \frac{\theta}{\rho} \left( 1 +
      \sum\limits_{m=1}^{\infty} \frac{\gamma^m}{(1+\rho)_m} \right),\\
    \mathbb E[D^{(n)}] &= \frac{\theta}{1 + \rho} \left( 1 +
      \sum\limits_{m=1}^{\infty} \frac{\gamma^m}{(2+\rho)_m } \right),\\
    \mathbb E[G^{(n)}] &= \theta \sum_{k=0}^{n-1}\frac{1}{k+\rho} +
    \theta \sum_{m=1}^\infty \frac{\gamma^{m}}{m}
    \Big(\frac{1}{(\rho)_m} - \frac{1}{(n+\rho)_m}\Big).
  \end{align}
\end{corollary}

\bigskip

\ISIsubtitleA{Discussion} %
We introduced the Wright--Fisher model for
bacterial genomes with horizontal gene flow, measured by the parameter
$\gamma$. Since the corresponding model without horizontal gene flow
was considered in \cite{BaumdickerHessPfaffelhuber2010}, we note that
Theorem~\ref{T1} and Corollary~\ref{C1} imply that all sample
statistics are continuous at $\gamma=0$. 

\begin{figure}
  \begin{center}
    \includegraphics[width=4in]{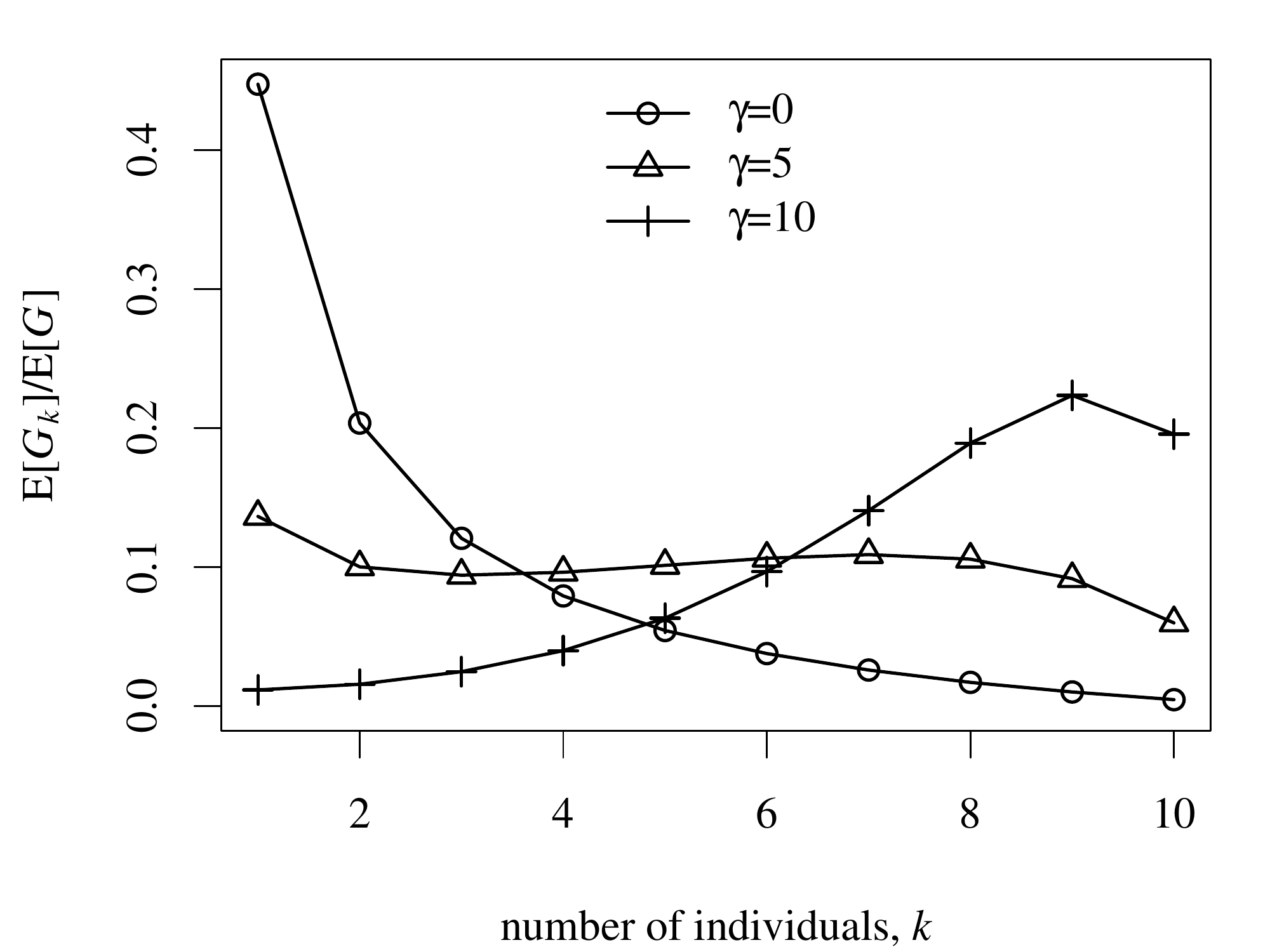}
    \caption{\label{fig2} The expected gene frequency spectrum is highly
      dependent of $\gamma$, the rate of horizontal gene flow. For high
      values of $\gamma$, most genes are in high frequency, leading to a
      closed pangenome. We use $\rho=2$ in the figure.}
  \end{center}
\end{figure}

Recently, the concepts of \emph{open} and \emph{closed} pangenomes
were introduced \citep{Medini2005}. If, after sequencing a finite
number of genomes, all genes present in the population are found, one
speaks of a closed pangenome. If new genes are found even after
sequencing many cells, the pangenome is \emph{open}. It is not hard to
see that high values of $\gamma$ imply that most genes are in
high-frequency. In other words, sequencing a new individual hardly leads
to new genes which were not seen before. This impact of openness and
closedness of the pangenome can as well be seen from
Figure~\ref{fig2}.

~

Consider the diffusion~\eqref{eq:SDE}, which describes the approximate
frequency of individuals carrying one specific gene $u\in I$.
Usually, the $\tfrac \gamma 2 X(1-X)$-term appears in population
genetic models only due to a selective force (see e.g.\
\citealp{Kimura1964,Ewens2004,Durrett2008}). In the present setting,
it appears because horizontal gene flow increases the frequency of the
gene by a rate which is proportional to the number of possible
donor/acceptor-pairs of individuals.

Due to the close connection of horizontal gene transfer with selective
models, a comparison to recent work is appropriate. In particular, the
theory for the frequency spectrum in selective models with
irreversible mutations is carried out in
\cite{Fisher1930,Wright1938,Kimura1964,Kimura1969}. Additionally,
\cite{SawyerHartl1992} developed a Poisson Random Field model for
selective sites. (Extensions were e.g.\ given in
\citealp{WilliamsonEtAl2005}.) They assume that a large set of
unlinked loci is under selection. As a result, they obtain predictions
for the number of alleles present in a subset $k$ out of $n$ of
individuals. However, since their loci are unlinked, the random
processes are completely independent for the different processes. This
is in contrast to our approach where the reproduction within the
Wright--Fisher model affects all genes in the same way, and the
horizontal gene transfer only affects single genes.  Moreover, our
model is reversible in the sense that present genes may as well be
lost (but not reintroduced). Since it has been shown in
\cite{BaumdickerHessPfaffelhuber2010} that discontinuities at $\rho=0$
(no gene loss) arise, it is not straight-forward to use these
classical results in the present setting.

\bigskip

\ISIsubtitleA{Simulations} %
We are interested in patterns of
presence/absence of genes in a sample of size $n$ in an equilibrium
situation. If presence/absence of a gene would be independent of the
state of the other genes, the gene frequency spectrum could be
simulated by independent copies of the diffusion $X$, given in
\eqref{eq:SDE}. However, all genes are inherited along the same
lineages, so the frequency of two different genes depend on each
other; see Figure \ref{fig4}. For $\gamma = 0$ the composition of the
accessory genome of $n$ individuals can be simulated backwards in time
using the coalescent. As, for $\gamma = 0$, genomes only depend on
events along the ancestral lineages of the sample this is very
efficient \citep{hudsonms}. In the case $\gamma>0$, we simulate the
accessory genome forward in time. Therefore, we have to consider all
individuals of the population, as each of them might influence the
accessory genome of the sample of size $n$ by events of horizontal
gene transfer. Unfortunately the size of the population has to be much
larger than $n$ to obtain values close to the large population limit
results. Thus the forward simulations for $\gamma >0$ are much slower
than backward simulations for $\gamma=0$.

\begin{figure}
  \begin{center}
    \includegraphics[width=3in]{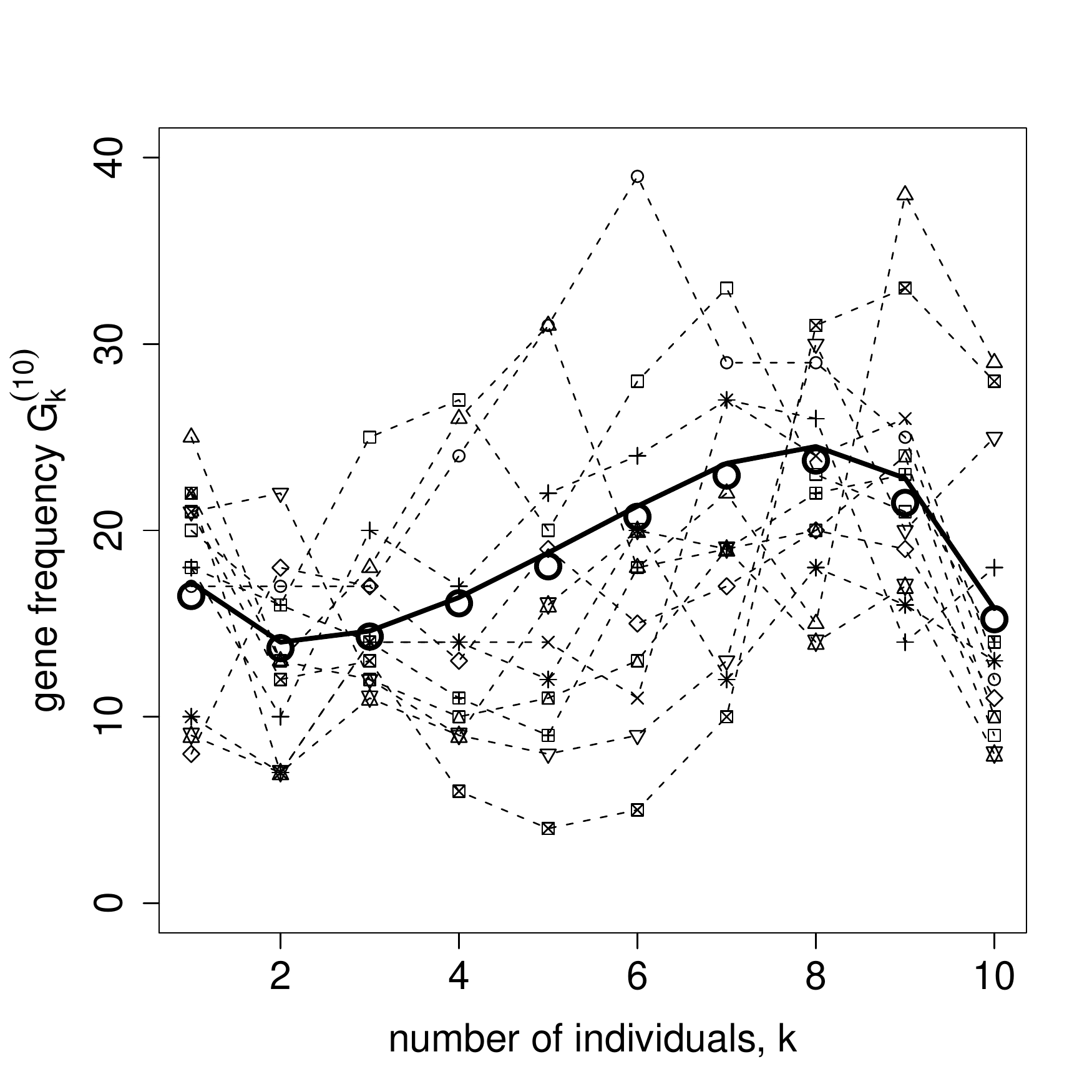}
    \caption{\label{fig5} The expected gene frequency spectrum for
      $n=10\text{, } \theta = 10\text{, } \gamma = 6$ and $\rho = 2$
      is shown as a solid black line.  For twelve different
      simulations with $N=500$ individuals the gene frequency spectrum
      for $n=10$ randomly chosen individuals is shown (dashed lines).
      The mean of 1000 simulations (black circles) is close to the
      results of Theorem \ref{T1}.}
  \end{center}
\end{figure}

\begin{figure}
  \begin{center}
    \includegraphics[width=6in]{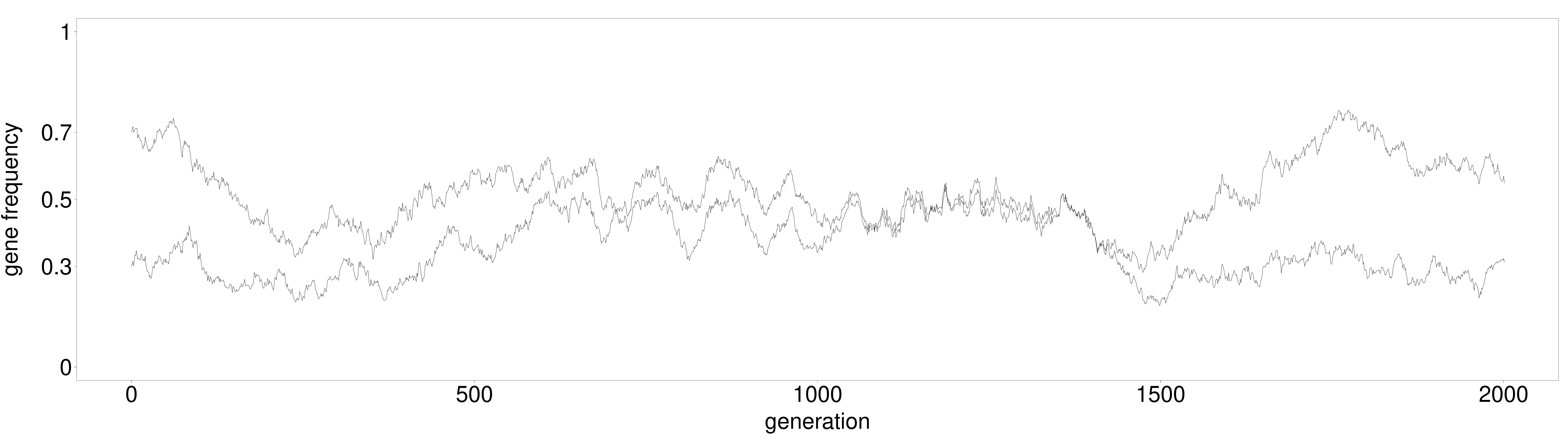}
    \caption{\label{fig4} The frequencies for two different genes in a
      population of size 2000 are shown. Here $\rho = 0.2$ and $\gamma
      = 1$.  At time zero initially 600 individuals carry gene 1, 800
      individuals carry both, gene 1 and gene 2 and 600 individuals
      carry none of the two genes. The frequencies are not independent
      as both genes depend on the same underlying ancestral
      lineages. Since gene loss and horizontal gene transfer events
      occur independently for each gene, they can weaken the
      dependency of the two frequency paths.}
  \end{center}
\end{figure}

Theorem \ref{T1} gives the expected sizes of the gene frequency
spectrum.  For $\gamma = 0$ it is known that the gene frequency
spectrum highly depends on the underlying genealogy. E.g. if the
genealogy separates the $n$ individuals into two groups, one of size
$k$ and one of size $n-k$, then $G_k^{(n)}$ and $G_{n-k}^{(n)}$
increase. The same is true for simulated data with $\gamma > 0$, see
Figure \ref{fig5}.

\bigskip

\ISIsubtitleA{Proof of Theorem~\ref{T1} and Corollary~\ref{C1}} %
We
consider the diffusion~\eqref{eq:SDE} with infinitesimal mean and
variance
\begin{align}\notag
  b(x) = -\tfrac 12 \rho x + \tfrac 12 \gamma x(1-x), \qquad a(x) =
  x(1-x).
\end{align}
The Green function for the diffusion, measuring the time the
diffusion, i.e. a gene, spends in frequency $x$ until eventual loss,
if the current frequency is $\delta$, is given
by 
\begin{align}\notag
  G(\delta,x) = 2 \frac{\phi(\delta)}{a(x)\psi(x)},
\end{align}
where
\begin{align*}
  \psi(y) &:= \exp\left( -2 \int_0^y \frac{b(z)}{a(z)} dz \right) = (1-y)^{1-\rho} e^{-\gamma y},\\
  \phi(x) &:= \int_0^x \psi(y) dy.
\end{align*}
Following \cite{Durrett2008}, we introduce new genes in frequency
$\delta$ at rate $\frac{\theta}{2} \frac{1}{\phi(\delta)}$ in a
consistent way. That is, the gene raises in frequency to
$\varepsilon>\delta$ with probability
$\frac{\phi(\delta)}{\phi(\epsilon)}$. Hence the number of genes in
frequency $x$ is Poisson with mean
\begin{align}\notag
  \frac{\theta}{2} \frac{1}{\phi(\delta)} G(\delta,x) = \theta
  \frac{e^{\gamma x}}{x(1-x)^{1-\rho}}.
\end{align}
The gene frequency spectrum is now given by
\begin{align}\notag
  \mathbb E[ G_k^{(n)} ] &= \binom{n}{k} \int_0^1 \theta
  \frac{e^{\gamma x}}{x(1-x)^{1-\rho}} x^k (1-x)^{n-k} dx \\ \notag &=
  \binom{n}{k} \theta \int_0^1 e^{\gamma x} x^{k-1} (1-x)^{n-k-1+\rho}
  dx \\ \notag &= \theta \binom{n}{k} (k-1)!
  \frac{\Gamma(n-k+\rho)}{\Gamma(n+\rho)} {}_1F_1(k;n+\rho;\gamma)
\end{align}
where ${}_1F_1(k;n+\rho;\gamma) = 1 + \sum\limits_{m=1}^{\infty}
\frac{(k)_m \gamma^m}{(n+\rho)_m m!}$ is a hypergeometric funtion and
$ (a)_b := a (a+1) \cdots (a+b-1).$ Given the gene frequency spectrum,
it is now easy to compute first moments of $A$, $D$ and $G$ (see
Corollary~\ref{C1}) by using
\begin{align*}
  \mathbb E[A] & = \mathbb E[G_1^{(1)}], \qquad \qquad \mathbb E[D] =
  \mathbb E[G_1^{(2)}], \\
  \mathbb E[G] & = \sum_{k=1}^n \tfrac 1k \mathbb E[G_1^{(k)}] =
  \sum_{k=1}^n \frac{\theta}{k} \frac{k}{k-1+\rho}\sum_{m=0}^\infty
  \frac{\gamma^m}{(k+\rho)_m} = \theta \sum_{m=0}^\infty {\gamma^{m}}
  \sum_{k=0}^{n-1} \frac{1}{(k+\rho)_{m+1}} \\ & = \theta
  \sum_{k=0}^{n-1}\frac{1}{k+\rho} + \theta\sum_{m=1}^\infty
  \frac{\gamma^{m}}{m} \sum_{k=0}^{n-1}\Big( \frac{1}{(k+\rho)_{m}} -
  \frac{1}{(k+1+\rho)_{m}}\Big) \\ & =\theta
  \sum_{k=0}^{n-1}\frac{1}{k+\rho} + \theta\sum_{m=1}^\infty
  \frac{\gamma^{m}}{m} \Big(\frac{1}{(\rho)_m} -
  \frac{1}{(n+\rho)_m}\Big).
\end{align*}

\bibliographystyle{chicago}
\bibliography{bacteria}

\end{document}